\documentclass{optica-article}

\journal{opticajournal} 

\articletype{Research Article}
\usepackage{lipsum}
\usepackage{lineno}

\makeatletter

\newcommand{\scriptveryshortarrow}[1][3pt]{{%
    \hbox{\rule[\scriptratio\dimexpr\fontdimen22\textfont2-.2pt\relax]
               {\scriptratio\dimexpr#1\relax}{\scriptratio\dimexpr.4pt\relax}}%
   \mkern-4mu\hbox{\let\f@size\sf@size\usefont{U}{lasy}{m}{n}\symbol{41}}}}

\makeatother

\usepackage{lipsum}  
\usepackage{siunitx}
\DeclareSIUnit \cycle {cyc}
\DeclareSIUnit{\belmilliwatt}{Bm}
\DeclareSIUnit{\dBm}{\deci\belmilliwatt}

\newcommand*\mean[1]{\overline{#1}}

\begin{document}

\title{Rapidly Tunable Synthetic Wavelength Ranging with an RFSoC}
\author{Shawn M. P. McSorley,\authormark{1, *} Benjamin P. Dix-Matthews,\authormark{1}, Andrew M. Lance,\authormark{1} David R. Gozzard,\authormark{1} and Sascha W. Schediwy \authormark{1}}

\address{\authormark{1}International Centre for Radio Astronomy Research, The University of Western Australia, Crawley, WA 6009, Australia}
\email{\authormark{*}Corresponding author: shawn.mcsorley@research.uwa.edu.au} 


\begin{abstract*}
  Measurements of optical range and time-of-flight are crucial for a variety of high-precision technologies. Competitive optical measurement techniques have been developed that balance precision with accuracy and system complexity. Here, we present a continuous-wave synthetic wavelength interferometry technique that employs digitally tunable electro-optic frequency combs. With a software-defined radio, our approach can dynamically sweep the synthetic wavelength and measure absolute optical range. We demonstrate this digital approach over a free-space optical delay line of \qty{1}{\meter} and over an \qty{40}{\kilo\meter} fiber link. The best obtained precision over the delay line is better than \qty{60}{\nano\meter} (\qty{0.2}{\femto\second}). Through a \qty{40}{\kilo\meter} fiber spool, this precision degrades to \qty{15}{\micro\meter} (\qty{50}{\femto\second}), which is a fractional error on the order of \qty{2d-10}{\meter/\meter}. Our design is simple to implement, and only relies on continuous-wave interference, decreasing system complexity.
\end{abstract*}
\section{Introduction}
The measurement of optical range, and optical time-of-flight, has had accelerated impact in the applications of time and frequency transfer \cite{sinclair2019, caldwell2023, caldwell2024, chen2024, olson2025, yu2024, bergeron2019, khader2018}, light detection and ranging, \cite{caldwell2022, camenzind2025, camenzind2022, sambridge2021, spollard2021, wang2022, barber2010, staffas2024, wu2024} and spectroscopy \cite{wong2020, wong2023, chan2025, han2024}. In applications requiring both precision and accuracy, continuous-wave optical measurement techniques, such as randomly modulated continuous wave \cite{wang2022, sambridge2021, spollard2021} and frequency modulated continuous wave, \cite{barber2010, staffas2024, wu2024} have been developed, and these avoid the ambiguity of the optical carrier at the cost of precision and complexity. Alternatively, optical frequency combs have enabled full interferometric precision with dual-comb implementations such as the time programmable frequency comb \cite{caldwell2022, caldwell2023, caldwell2024}, pulsed time-of-flight with linear optical sampling, \cite{coddington2009,sinclair2019, carlson2018, martin2022, li2022} and synthetic-wavelength interferometry (SWI) \cite{lay2003,zhi2025,yang2015, zhao2018, weimann2018, li2025,  xie2023,  zhu2018, mcsorley2025}.

The interference of two frequency combs using linear optical sampling is a well established technique. It has been used to provide interferometric precision over \qty{100}{\kilo\meter}+ free-space baselines \cite{shen2022, han2024} and with platforms under motion \cite{sinclair2019, bergeron2019, camenzind2025}. Development of the application of optical sampling over long-haul fiber networks is also underway \cite{chen2024, olson2025, yu2024}, addressing dispersion impairments of the pulse coherence between either frequency comb. Optical sampling requires strict phase coherence between the combs, which excludes its application from pre-existing optical fiber networks that do not compensate for pulse dispersion. Furthermore, it is limited by its poor photon efficiency, limited to optical return powers of a few nanowatts \cite{caldwell2023}.

The photon inefficiency has been addressed by the time programmable frequency comb \cite{caldwell2022, caldwell2023, caldwell2024}, capable of femtosecond performance with picowatt return powers. However, when balancing system complexity and absolute precision, SWI seeded by continuous-wave sources can be favorable. There has been increased interest in electro-optic comb generation for SWI \cite{yang2015, zhao2018, weimann2018, li2025, xie2023, zhu2018, mcsorley2025}. In part due to its simple, yet versatile, tunability of comb lines, and its natural coherence provided by the continuous-wave source. Electro-optic combs provide a competitive combination of non-ambiguous range, precision, and update rate, which can be optimized by the electrical source driving the electro-optic modulator (EOM). 

SWI is realized by the combination of distinct optical frequencies, where a synthetic wavelength is constructed from unique carrier phase measurements at separate optical frequencies.  Heterodyne beat note techniques have been employed \cite{lay2003, yang2015, xie2023, mcsorley2025, weimann2018, zhu2018} to obtain the necessary optical carrier phases from the interference of two detuned and offset electro-optic combs. These heterodyne techniques enable high precision, through driving the modulators with microwave frequencies, while only requiring radio frequency bandwidth detection electronics. By tuning the line spacing of each comb, a dynamic ambiguous range can be obtained, at the expense of synthetic wavelength precision. To overcome this compromise, frequency hopping of the comb line spacing has been investigated in \cite{xie2023} and \cite{mcsorley2025}.

The heterodyne approach also provides several advantages when compared to linear optical sampling. Firstly, it is compatible with existing free-space optical frequency transfer techniques \cite{gozzard2022, dix-matthews2021, mcsorley2025a, mcsorley2023}, which have also successfully been demonstrated over moving links \cite{dix-matthews2023, mcsorley2025a}. The performance of the continuous-wave approach, with only two dominant optical sidebands, is expected to have a superior quantum limit \cite{mcsorley2025, sambridge2023} when compared to the quantum limit of pulse-to-pulse interference \cite{caldwell2024, martin2022}. Lastly, the coherence requirements for continuous-wave interferometry are relaxed when compared to pulse techniques.

The advancement of digital signal processors has also accelerated the performance of the aforementioned techniques. Field-programmable gate arrays (FPGA) have been utilized in time-programmable combs \cite{caldwell2022}, in optical sampling \cite{sinclair2019} and in SWI \cite{xie2023, mcsorley2025, weimann2018}. Graphical processing units have also been employed to offload data processing for optical sampling, providing a more intuitive programming platform \cite{camenzind2025}. Regardless, FPGAs can provide flexibility, parallelism, and deterministic latency in signal processing tasks, which is favorable for high-speed optical signal processors. Even more beneficial is the integration of processors, in the form of a system-on-chip (SoC). An SoC provides the capability to offload data, at a lower sample rate than the FPGA, to the processor for post-processing or real time software control.

Our previous research demonstrated an all-digital SWI scheme \cite{mcsorley2025}, in which we employed an ADALM-Pluto (Zynq7010 SoC + AD9363 RF Front End) software-defined radio to digitally control the EOM on command, while implementing digital phase-locked loops (PLLs) in parallel on the FPGA to extract the phase of each heterodyne beat note. PLLs are capable of tracking optical phase without the need for $2\pi$ phase unwrapping, and they can be tuned for a variety of experimental conditions, including weak-light tracking \cite{sambridge2023} and for rapidly moving links \cite{dix-matthews2023, mcsorley2025a}. Continuous-wave PLL measurements are notoriously prone to ambiguities, particularly when loss-of-lock occurs. However, the combination of continuous-wave interferometry and SWI can remove this ambiguity. By using a software defined radio, the EOM can be dynamically, and rapidly, tuned as required to resolve successive ambiguities in the synthetic wavelength chain.

\begin{figure}
  \centering
  \includegraphics{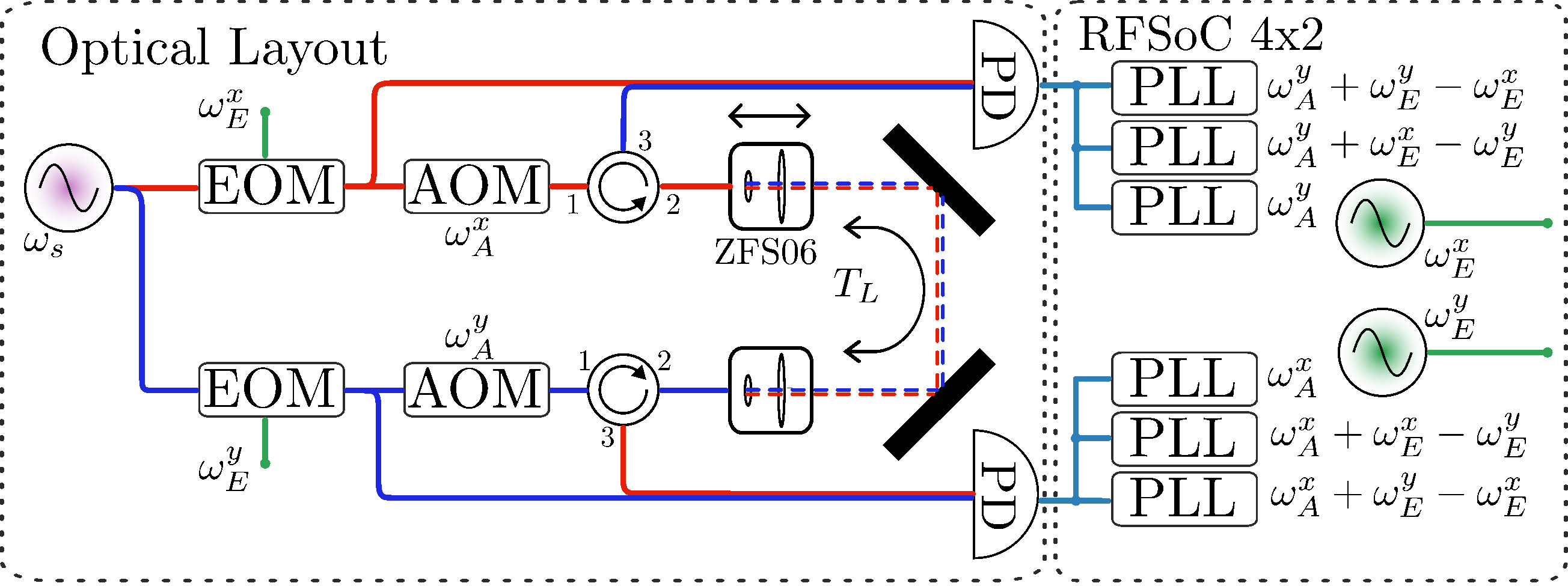}

  \caption{Schematic diagram showing the optical and radio-frequency layout for digital multi-wavelength optical absolute ranging. EOM, electro-optic modulator; AOM, acousto-optic modulator; PD, photodetector; PLL, phase-locked loop; $\omega_s$, optical source frequency; $\omega^x_E$ and $\omega^y_E$, EOM drive frequencies; $\omega^x_A$ and $\omega^y_A$, AOM drive frequencies; $T_{L}$, optical link delay mechanically actuated with a motorized collimator (ZFS06 stepper motor).}
  \label{fig:experimental_diagram}
\end{figure}

In this article, we improve upon our previous work by implementing our SWI scheme entirely in a radio-frequency SoC (RFSoC). We describe a digital synthetic-wavelength absolute ranging scheme, capable of tracking moving optical links over a demonstrated \qty{5}{\milli\meter} range. Our approach utilizes the RFSoC to perform both synthetic wavelength stepping and PLL phase tracking, in intervals as short as \qty{0.1}{\milli\second}. We experimentally demonstrate absolute ranging on a short \qty{1}{\meter} free-space delay line, and over a \qty{40}{\kilo\meter} fiber link. For the delay line, we obtain precision on the order of \qty{60}{\nano\meter} (\qty{0.2}{\femto\second}). For the \qty{40}{\kilo\meter} link, we achieve a best residual precision on the order of \qty{15}{\micro\meter} (\qty{50}{\femto\second}) when compared to the displacement of the optical carrier, which is a fractional error on the order of \qty{2d-10}{\meter/\meter}.  
\section{Experimental Method}
\subsection{Optical Design and Theory}

The experimental layout of our dynamic digital synthetic-wavelength ranging scheme is provided in Figure \ref{fig:experimental_diagram}. A laser, with an optical frequency denoted by $\omega_s$, seeds two identical optical transceivers. Optical sidebands are generated in both transceivers via electro-optic phase modulators (EOM), with a spacing given by the electrical drive frequencies denoted $\omega^x_E$ and $\omega^y_E$. The instantaneous phase of the optical sidebands, at the output of the EOMs, can be described as,
\begin{align}
  \phi^i_k(t) &= [\omega_s + k \omega^i_E]t + \delta\phi^i_k(t),\text{ and}\\
  \delta\phi^i_k(t) &= k \delta\phi^i_E(t) + \delta\phi_s(t). 
\end{align}
Here $t$ is time, $i=x,y$ represents the transceiver, $k=-1,0,1$ represents the lower sideband, the carrier and the upper sideband respectively, $\delta\phi_s(t)$ is the modeled phase noise of the laser, and $\delta\phi^i_E(t)$ is the modeled phase noise of the radio-frequency driving the EOMs. 

 The output of the EOM provides both a local reference and an outgoing signal. Acousto-optic modulators (AOM) provide a frequency shift for the outgoing signals with frequency $\omega^x_A$ and $\omega^y_A$, for transceiver x and y respectively. The phase of the outgoing signals from transceiver x and y can respectively be described by:
\begin{align}
  \phi^{i\rightarrow}_k(t) &= [\omega_s + k \omega^i_E + \omega^i_A]t + \delta\phi^{i\rightarrow}_k(t),
\end{align}
where $\delta\phi^{i\rightarrow}_k(t)=\delta\phi^i_k(t)+\delta\phi^i_A(t)$ and $\delta\phi^i_A(t)$ is the phase noise of the AOMs RF drive signal.

Our experimental demonstration includes a free-space optical delay line, denoted by $T_{L}=n_{a}L/c$, where $n_{a}$ is the refractive index of air and $L$ is the free-space distance. The delay line is constructed with a collimator mounted to a motorized translation stage. To test the absolute ranging performance over a long baseline, we optionally include a long fiber link of length $L_f$. The optical delay of this fiber link can be lumped into $T_{L}$ as $T_{L}=n_{a}L/c+n_fL_f/c$, where $n_f$ is the refractive index of the fiber. The phase of the incoming signals to transceiver x and y can respectively be described by:
\begin{align}
  \phi^{x\leftarrow}_k(t)=\phi^{y\rightarrow}_k(t-T_{L})=&(\omega_s + k \omega^y_E + \omega^y_A)[t-T_{L}] + \delta\phi^{y\rightarrow}_k(t-T_{L}),\text{ and}\\
  \phi^{y\leftarrow}_k(t)=\phi^{x\rightarrow}_k(t-T_{L})=&(\omega_s + k \omega^x_E + \omega^x_A)[t-T_{L}] + \delta\phi^{x\rightarrow}_k(t-T_{L}).
\end{align}
\begin{figure}

  \centering
  \includegraphics{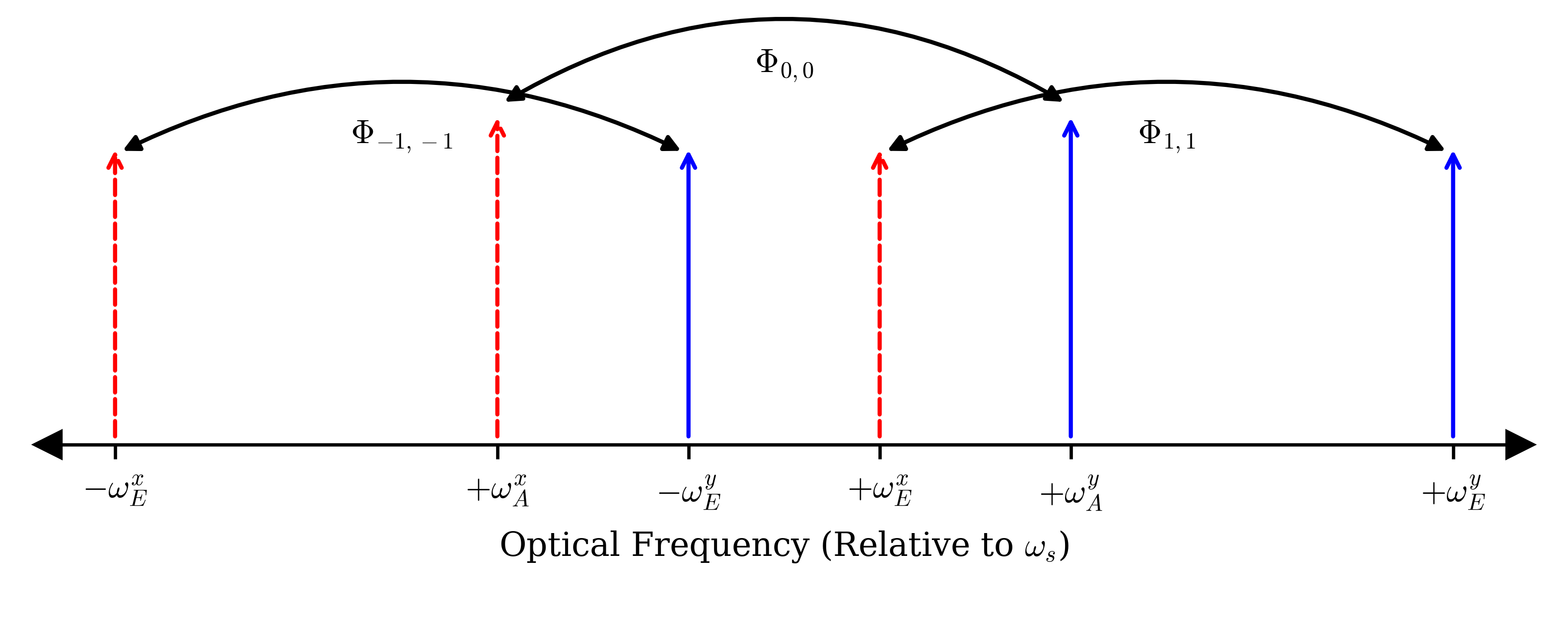}
  \caption{Simplified illustration of the beat process at either optical transceiver. The optical signal from transceiver x is shown in dashed red, and the optical signal from transceiver y is shown in blue. Each transceiver tracks three beatnotes. }
  \label{fig:sideband}
\end{figure}

At either transceiver, we consider three beat notes, $\Phi^i_{k,k}=\phi^{i\leftarrow}_k-\phi^i_k$. A simplified illustration of the beating process is provided in Fig \ref{fig:sideband}. We do not consider higher order sidebands. To ensure spectral separation of each beat note, we require $\omega^x_E \neq \omega^y_E$. 
The instantaneous phase of each beat note is then,  
\begin{align}
  \Phi^x_{k,k}(t)&=[k\{\omega^y_E-\omega^x_E\}t+\omega^y_A]t-[k\omega^y_E+\omega^y_A+\omega_s]T_{L} +\delta\phi^{y\rightarrow}_k(t-T_{L})-\delta\phi^x_k(t)+N^x_k, \text{ and}\\
  \Phi^y_{k,k}(t)&=[k\{\omega^x_E-\omega^y_E\}t+\omega^x_A]t-[k\omega^x_E+\omega^x_A+\omega_s]T_{L} +\delta\phi^{x\rightarrow}_k(t-T_{L})-\delta\phi^y_k(t)+N^y_k
\end{align}

Here, we introduce $N^x_{k}$ and $N^y_{k}$ which are unknown integer cycles of phase in each PLL measurement. The PLLs are configured to track phase variations around the nominal frequency of each sideband. These phase measurement can be combined to obtain
\begin{align}
  \Phi_{\lambda'}&=\sum_{i=x,y}(\Phi^i_{1,1}-\Phi^i_{-1,-1}) = -2[\omega^x_E+\omega^y_E]T_{L}+\delta\phi(t)+ N_{\lambda'}, \label{eqn:sideband}
\end{align}
providing a synthetic wavelength of $\lambda'=c/2(\omega^x_E+\omega^y_E)$. Here, $N_{\lambda'}$ is the lumped unknown integer number of cycles and $\delta\phi$ is the net phase noise contribution, where,
\begin{equation}
  \delta\phi(t)=2[\delta\phi^x_E(t-T_{L}) + \delta\phi^y_E(t-T_{L}) - \delta\phi^x_E(t) - \delta\phi^y_E(t)]. \label{eqn:EOM_NOISE} 
\end{equation}
The immediate benefits of the synthetic wavelength technique are the net cancellation of laser phase noise and any phase noise introduced by the AOMs, and suppression of the EOM radio-frequency noise governed by $T_{L}$.

An important consideration is that the SNR of each beat note will be limited by the shot noise of the strong local oscillator. This SNR implies that each phase measurement will have an additive shot noise term \cite{sambridge2023}, with a phase power spectral density (PSD) given by,
\begin{equation}
  S^i_k=\frac{1}{J^2_k(KV_x)J^2_k(KV_y)} \frac{\hbar c}{ 2 \pi \eta \lambda P_i}.\quad [\si{\cycle\squared\per\hertz}]
\end{equation}
Here, $P_i$ is the received optical power from the link at either transceiver's photodetector, $\lambda$ is the optical wavelength, $\eta$ is the quantum efficiency, $J$ is a Bessel function of the first kind, $K$ is the EOM modulation index and $V_i$ is the drive voltage of the EOM at either transceiver. The EOMs are driven to balance the power between the carrier and inner most sidebands, and this results in an 11~dB shot noise penalty on each beat note. However, as shot noise is uncorrelated in each phase measurement, the total phase shot noise contribution will be

\begin{equation}    
  S=\frac{2}{J^2_1(KV_x)J^2_1(KV_y)} \frac{\hbar c}{ 2 \pi \eta \lambda P_x}+\frac{2}{J^2_1(KV_x)J^2_1(KV_y)} \frac{\hbar c}{ 2 \pi \eta \lambda P_y}. \quad [\si{\cycle\squared\per\hertz}] \label{eqn:shot_noise}
\end{equation}

The synthetic wavelength in Equation \ref{eqn:sideband} can be tuned by changing the EOM drive frequencies. We denote a particular synthetic wavelength by $\lambda_j$, where $j$ corresponds to an EOM configuration, providing the average link estimates,
\begin{align}
  \mean{T}_{L}&=\frac{\mean{\Phi}_{\lambda_j}+N_{\lambda_j}}{2(\omega^x_E+\omega^y_E)}\\
  \mean{L} &= \lambda_j\mean{\Phi}_{\lambda_j}+\lambda_j N_{\lambda_j}, 
\end{align}
where $N_{\lambda_j}$ is an integer cycle ambiguity. For a wavelength change, $\lambda_{j+1}$, that is faster than variations in $L$, we assume that
\begin{equation}
\lambda_j\mean{ \Phi}_{\lambda_j}+\lambda_j N_{\lambda_j}=\lambda_{j+1}\mean{ \Phi}_{\lambda_{j+1}}+\lambda_{j+1}N_{\lambda_{j+1}}.
\end{equation}
If the precision provided by $\Phi_{\lambda_j}$ satisfies $\sqrt{\operatorname{Var}(\lambda_j \Phi_{\lambda_j})}  < \lambda_{j+1}/4$, we can progressively determine integer cycle ambiguities with
\begin{equation}
  N_{\lambda_{j+1}}=\frac{\lambda_j\mean{ \Phi}_{\lambda_j}+\lambda_j N_{\lambda_j}-\lambda_{j+1}\mean{ \Phi}_{\lambda_{j+1}}}{\lambda_{j+1}}. \label{eqn:integerCyc}
\end{equation}
For short optical links, where the microwave frequency noise is largely suppressed and other fiber effects are negligible, this precision is set by Equation \ref{eqn:shot_noise}. This precision follows an Allan deviation (ADEV), in meters, of $\sigma_L(\tau)=\lambda'\sqrt{S/2\tau}$, for an integration time $\tau$ in seconds.

Furthermore, the carrier phases can also be combined to provide an independent displacement measurement with interferometric sensitivity,
\begin{equation}
  \Phi_\lambda=\Phi^x_{0,0}+\Phi^y_{0,0}=-T_{L}[2\omega_s+\omega^y_A+\omega^x_A]+2\delta\phi_s(t-T_{L})-2\delta\phi_s(t). \label{eqn:carrier}
\end{equation}
Throughout the manuscript, we compare the synthetic link estimates obtained from $\Phi_{\lambda'}$ with the carrier displacement $\Phi_{\lambda}$.

\begin{figure}
  \centering
  \includegraphics{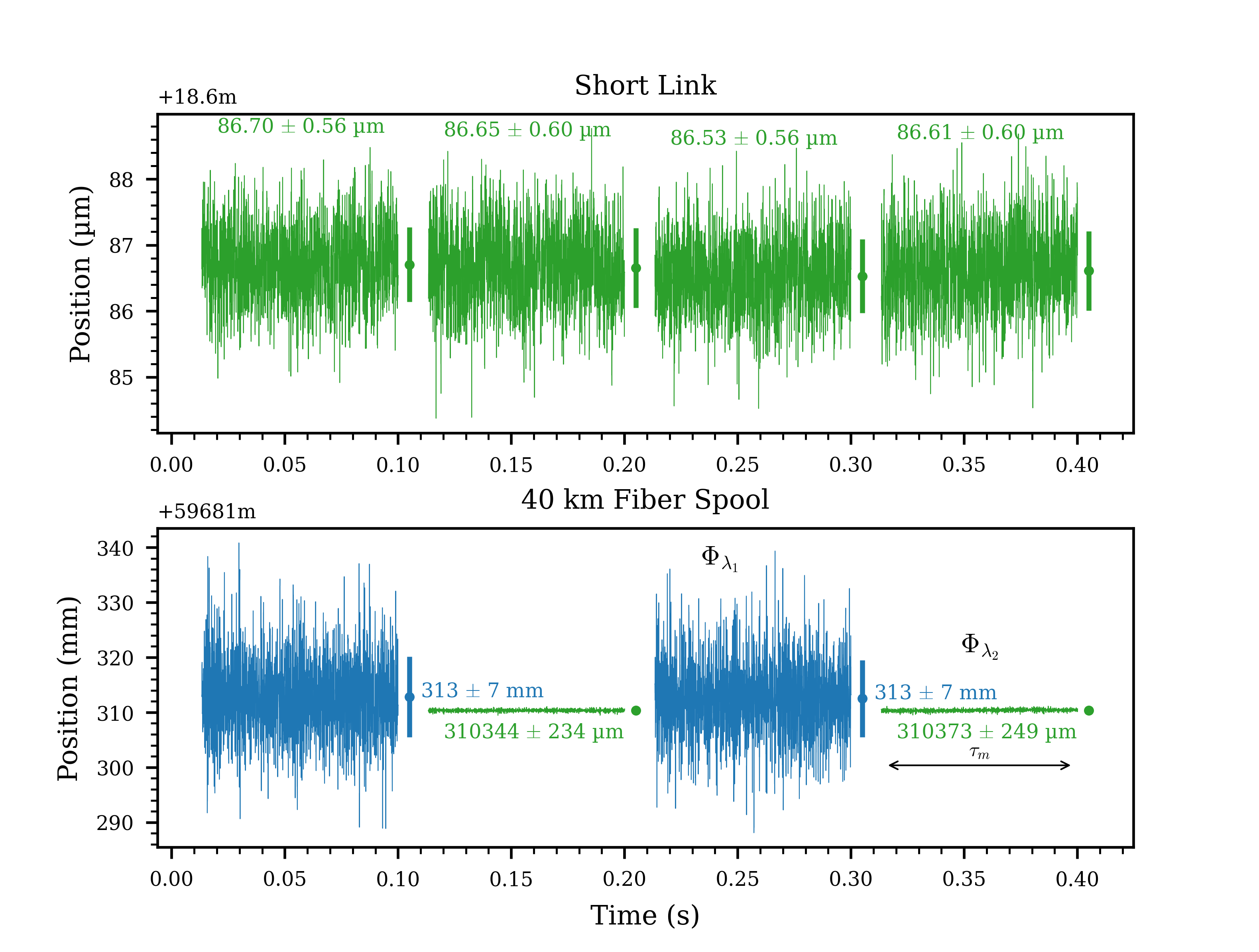}
  \caption{Data processing for digital multi-wavelength ranging. For the short link (top panel), the synthetic wavelength is repeatedly reset. For the \qty{40}{\kilo\meter} fiber link (bottom panel), the synthetic wavelength is repeatedly stepped. The synthetic phase of the wavelength steps, $\Phi_{\lambda_1}$ (blue) and $\Phi_{\lambda_2}$ (green), are repeatedly measured. Each measurement has an interval of $\tau_m$. The mean is taken from each time-series to form a link estimate. The mean and standard deviation are drawn as error bars next to each measurement interval. Equation \ref{eqn:integerCyc} is used to determine the integer cycle of each time series, relative to the \textit{a priori} link length.}
  \label{fig:steppingProcess_png}
\end{figure}

\subsection{Experimental Demonstration}
The experiment is seeded by a commercial off the shelf \qty{1550}{\nano\meter} laser (NKT X15, line width \qty{100}{\hertz}, power \qty{14}{\dBm}). A \qty{1}{\meter} optical delay line is created with a Thorlabs ZFS06 stepper motor, and two Thorlabs F240 collimators. The fiber optics used in the ranging system are polarization maintaining, with \qty{12}{\meter} of in-loop fiber. 


To provide a clear shot noise margin, the optical power of the reference arm in both transceivers was set to \qty{-18}{\deci\belmilliwatt}. The maximum return power into each transceiver was $P_x=\qty{-21}{\deci\belmilliwatt}$, and $P_y=\qty{-16}{\deci\belmilliwatt}$.

The EOM frequencies are generated by a Real Digital RFSoC 4$\times$2 SDR, featuring an RF Data Converter (RFDC) and Zynq Ultrascale+ FPGA. To generate tunable EOM drive frequencies, two numerically controlled oscillators (NCO), located within the RFDC, are digitally commanded from the FPGA. The RFDC is able to step the NCO within \qty{6}{\micro\second}. 

To measure the phase of each beatnote, six digital PLLs are implemented inside the FPGA. The PLL design is identical to that presented in \cite{mcsorley2023}. The benefits of the digital PLL approach includes rapid tunability, enabling the tracking of both dynamic and weak optical signals \cite{mcsorley2025a,mcsorley2025, sambridge2023}; phase data can be offloaded for post processing; and, for our absolute ranging approach, we can dynamically reset the PLLs at sub-\si{\kilo\hertz} rates.

To investigate the performance of our digital ranging system, we use two experimental configurations. The first, shown in Figure \ref{fig:experimental_diagram}, includes the aforementioned optical delay line which can provide link variations with a range up to \qty{5}{\milli\meter}. The second configuration includes a \qty{40}{\kilo\meter} SMF-28 fiber spool link. To dynamically track variations of the optical path, we program the FPGA to both step the synthetic wavelength and measure the synthetic phase.

For the short link, the RFDC is repeatedly commanded to step to an EOM frequency of \qty{4}{\giga\hertz}. This provides a constant synthetic wavelength of \qty{18}{\milli\meter}, which ensures a sufficient non-ambiguous range for the \qty{12}{\meter} of in-loop fiber. While the wavelength step is the same, both the EOM drive frequency and PLLs are reset each measurement. This frequency was within the second Nyquist zone of the RFSoC, requiring an analog bandpass filter to ensure sufficient spectral purity. 

For the \qty{40}{\kilo\meter} link, the RFDC is repeatedly commanded to step between \qty{100}{\mega\hertz} and \qty{2}{\giga\hertz}, providing synthetic wavelengths of \qty{750}{\milli\meter} and \qty{36}{\milli\meter} respectively. These steps were both chosen to be within the first Nyquist zone of the RFSoC. An example of the data processing for both links is provided in Fig. \ref{fig:steppingProcess_png}.

The PLL bandwidths are set to roughly $\qty{50}{\kilo\hertz}$. We vary the measurement intervals $\tau_m$, from \qty{0.1}{\milli\second} to \qty{10}{\milli\second}, and also compare to identical experimental measurements where the PLLs are not reset. This is done to demonstrate the ranging capability of the SWI system over time, where consecutive measurements are repeatedly taken in a single shot fashion.

The phase measurements recorded within the FPGA are offloaded to the CPU for processing. The integer wavelengths are successively determined for each time series using Eqn. \ref{eqn:integerCyc}. They are referenced to the \textit{a priori} link distance of $n_aL_{link}+n_fL_f\approx\qty{18}{\meter}$. A link estimate can be obtained from each cycle of wavelength steps, where the mean of the smallest synthetic wavelength time series provides a single data point.

\begin{figure}
  \centering
  \includegraphics{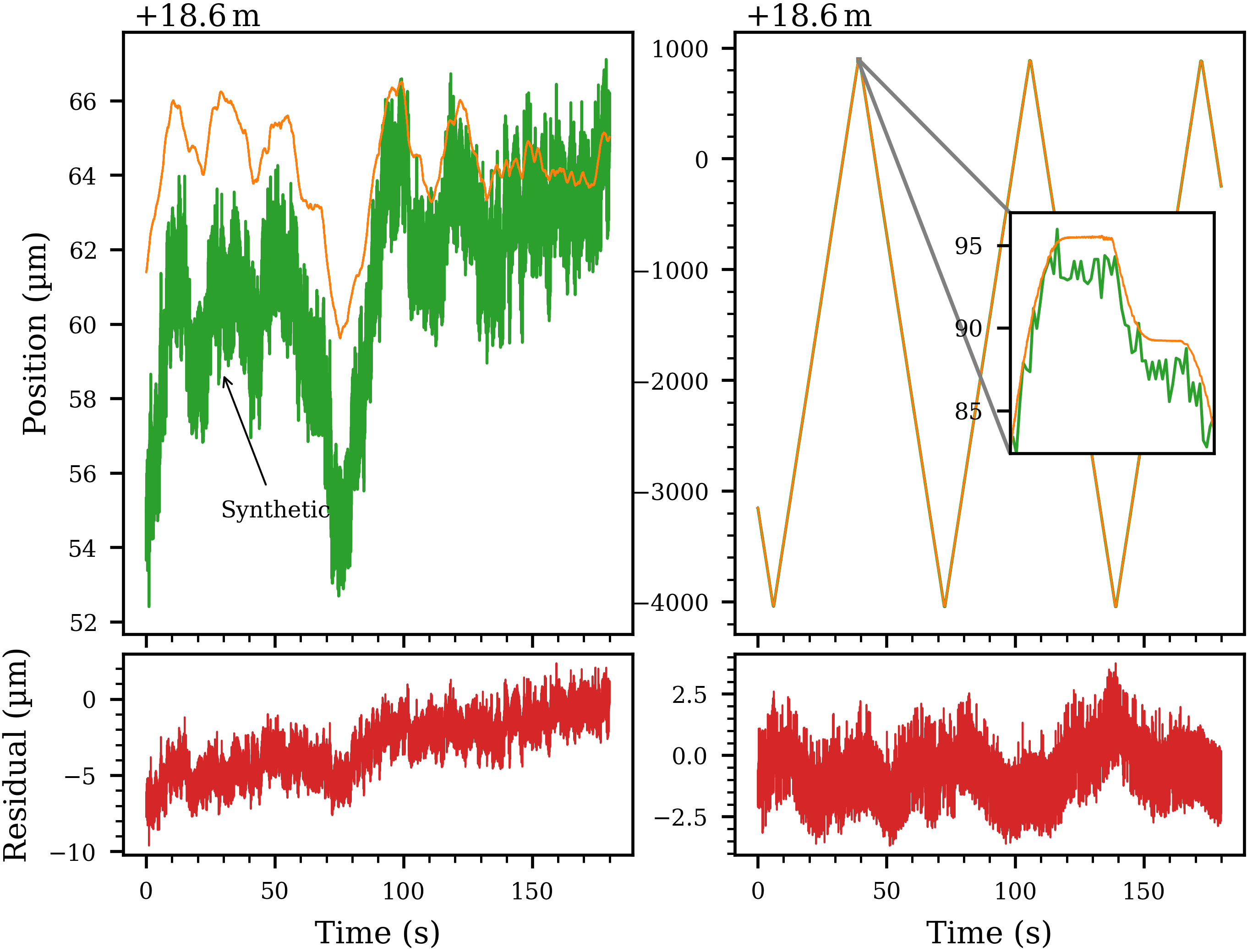}
  \caption{Absolute range estimates for the short link. The left panel shows the synthetic range estimate (green), the carrier displacement measurement (orange) and their difference residual (red) for a stationary optical link distance of \qty{18.6}{\meter}. The right panel shows the synthetic range estimate (green), the carrier displacement measurement (orange) and their difference residual (red) for the same optical link with a continuous \qty{5000}{\micro\meter} displacement.}
  \label{fig:results_fiber}
\end{figure}

\section{Experimental Results and Discussion}

To assess the performance of the absolute ranging system over long time intervals, we first record absolute range estimates of the short link with a stationary optical delay line, and then with continuous forward and backward movement over a \qty{5}{\milli\meter} range. These are shown in the left and right panels of Fig. \ref{fig:results_fiber} respectively. The synthetic distance estimates are provided in green, obtained using the process show in Fig. \ref{fig:steppingProcess_png}, with $\tau_m=\qty{10}{\milli\second}$. The \textit{a priori} knowledge of the fiber and optical delay line lengths is used to determine the initial range estimate. The carrier phase (orange), from Eqn.~\ref{eqn:carrier}, provides an interferometric displacement measurement, independent of the EOM phase modulation. The residual between our absolute range estimate and the carrier displacement measurement is shown in red. These measurements are repeated with a continuous phase measurement, where the PLLs are not reset, and a measurement interval of $\tau_m=\qty{0.1}{\milli\second}$.

The two measurements are well correlated as expected. For the stationary link, the disagreement between the two measurements drifts by up to \qty{6}{\micro\meter} over \qty{180}{\second}, with short term variations of roughly \qty{3}{\micro\meter} peak-to-peak. 

\begin{figure}
  \centering
  \includegraphics{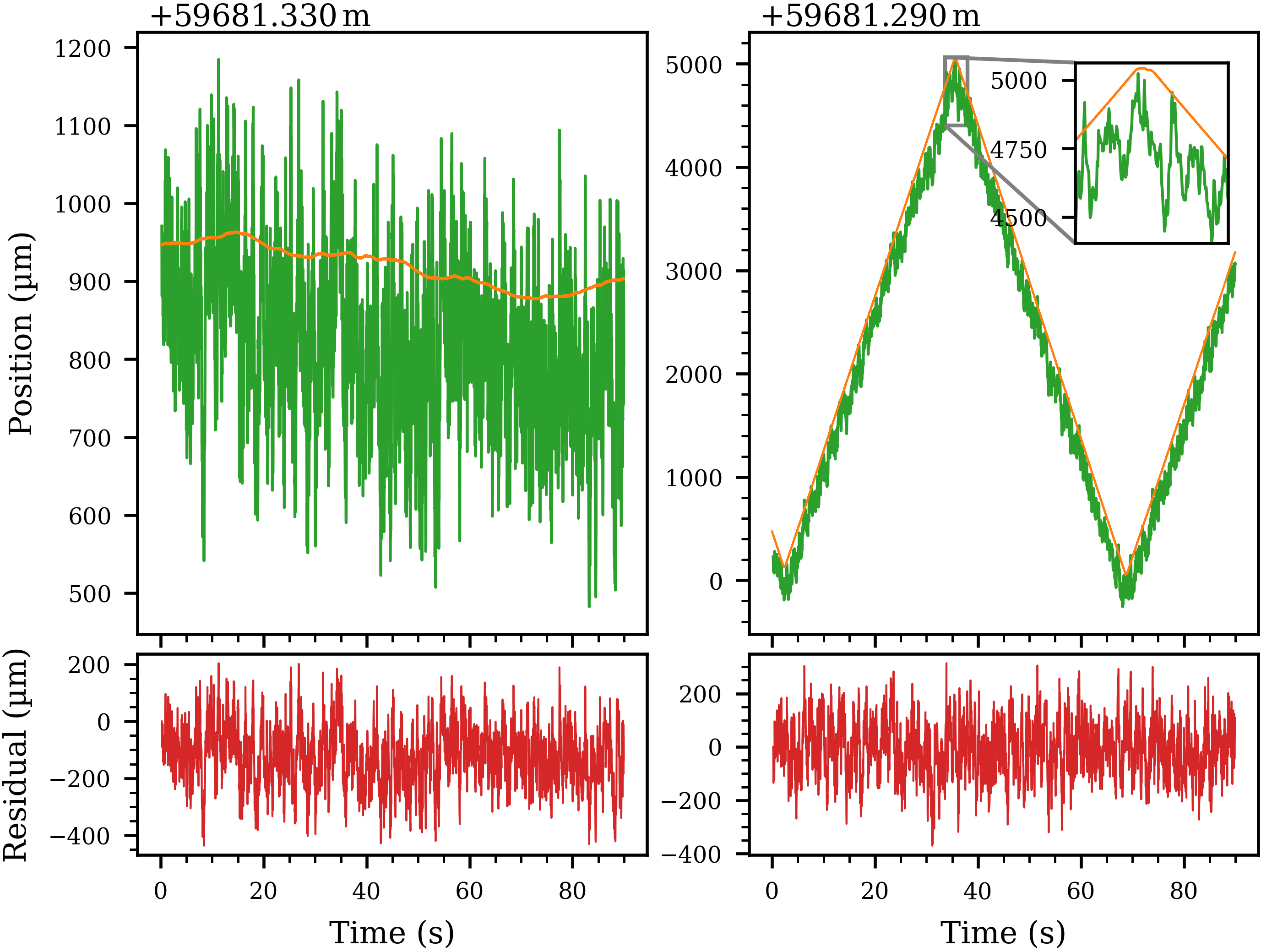}
  \caption{Absolute range estimates for the \qty{40}{\kilo\meter} fiber link. The left panel shows the synthetic range estimate (green), the carrier displacement measurement (orange) and their difference residual (red) for a stationary free-space optical link distance of \qty{59681}{\meter}. The right panel shows the synthetic range estimate (green), the carrier displacement measurement (orange) and their difference residual (red) for the same optical link with a continuous \qty{5000}{\micro\meter} displacement.}
  \label{fig:results_spool}
\end{figure}

The measurements are repeated for the \qty{40}{\kilo\meter} link, shown in Fig. \ref{fig:results_spool}. We estimate an initial fiber distance of \qty{40,682}{\meter} using a pseudo-random bit sequence and digital correlator, similar to that presented in \cite{khader2018}. This initial estimate is then used as an \textit{a priori} input into the data processing shown in Fig. \ref{fig:steppingProcess_png}. We see an increased noise floor in both measurements, with peak-to-peak fluctuations now on the order of \qty{400}{\micro\meter}. For the short link, we expect short term variations to be set by the shot noise limit in Eqn.~\ref{eqn:shot_noise}. There is no active amplification or polarization compensation for the \qty{40}{\kilo\meter} fiber link, which incurs roughly a \qty{20}{\deci\bel} penalty on each optical sideband. However, this power penalty does not account for the observed 100 times increase in short term fluctuations over the longer link. We also see no obvious impairments related to fiber dispersion on the \qty{}{\milli\meter} scale, however, further analysis is required to understand if it impacts the observed residual on the \qty{}{\micro\meter} scale. 

\begin{figure}
  \centering
  \includegraphics{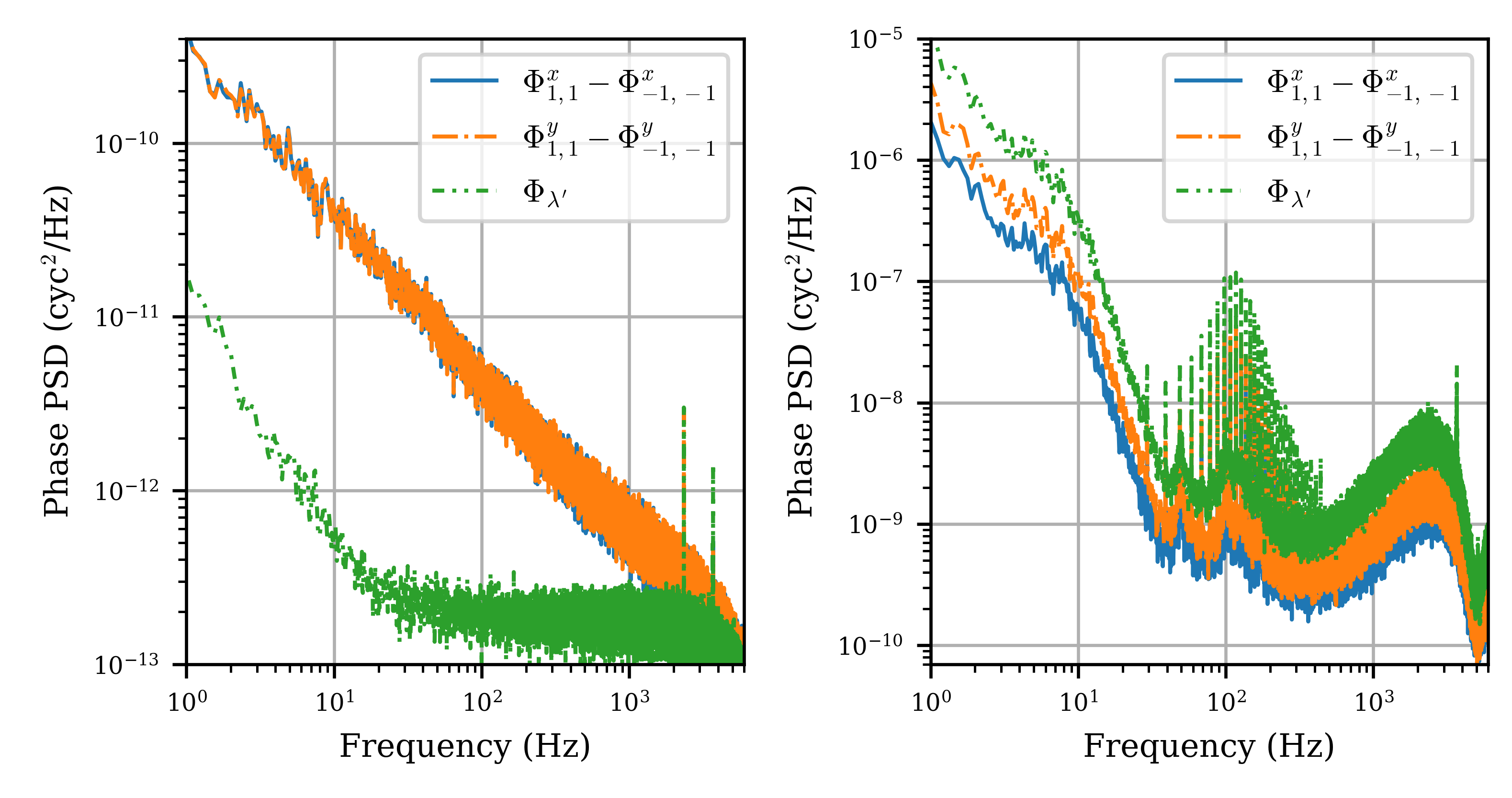}
  \caption{Phase power spectral densities (PSD) for sideband combinations in Equation~\ref{eqn:sideband}. The left panel shows the PSDs for the short link, while the right panel shows the PSDs for the \qty{40}{\kilo\meter} link. These are obtained from stationary $\Phi_{\lambda_2}$ time-series.}
  \label{fig:PSDs_png}
\end{figure}

To understand the noise properties over short-timescales, the phase power spectral density for both links is provided in Fig.~\ref{fig:PSDs_png}. Using a static link configuration, we calculate the PSD of each sideband combination in Equation ~\ref{eqn:sideband}, for both the short link (left panel) and \qty{40}{\kilo\meter} link (right panel). As expected, the short link is dominated by a white noise floor above \qty{20}{\hertz}, attributed to Equation~\ref{eqn:shot_noise}. However, the \qty{40}{\kilo\meter} link is clearly dominated by time-delayed phase noise between \qty{100}{\hertz} and the expected roll-off at \qty{5000}{\hertz}. This is instead attributed to Eqn. \ref{eqn:EOM_NOISE}, where we see the microwave phase noise is increased in the synthetic combination. 

Another consideration is the spectral purity of the EOM drive frequencies. To enable wavelength stepping on the long link, a bandpass filter could not be used. We have observed that spurious tones on the EOM can degrade the measurement performance. Despite this, we are able to resolve sub-\qty{}{\milli\meter} movement over \qty{40}{\kilo\meter} of optical fiber.

\begin{figure}
  \centering
  \includegraphics{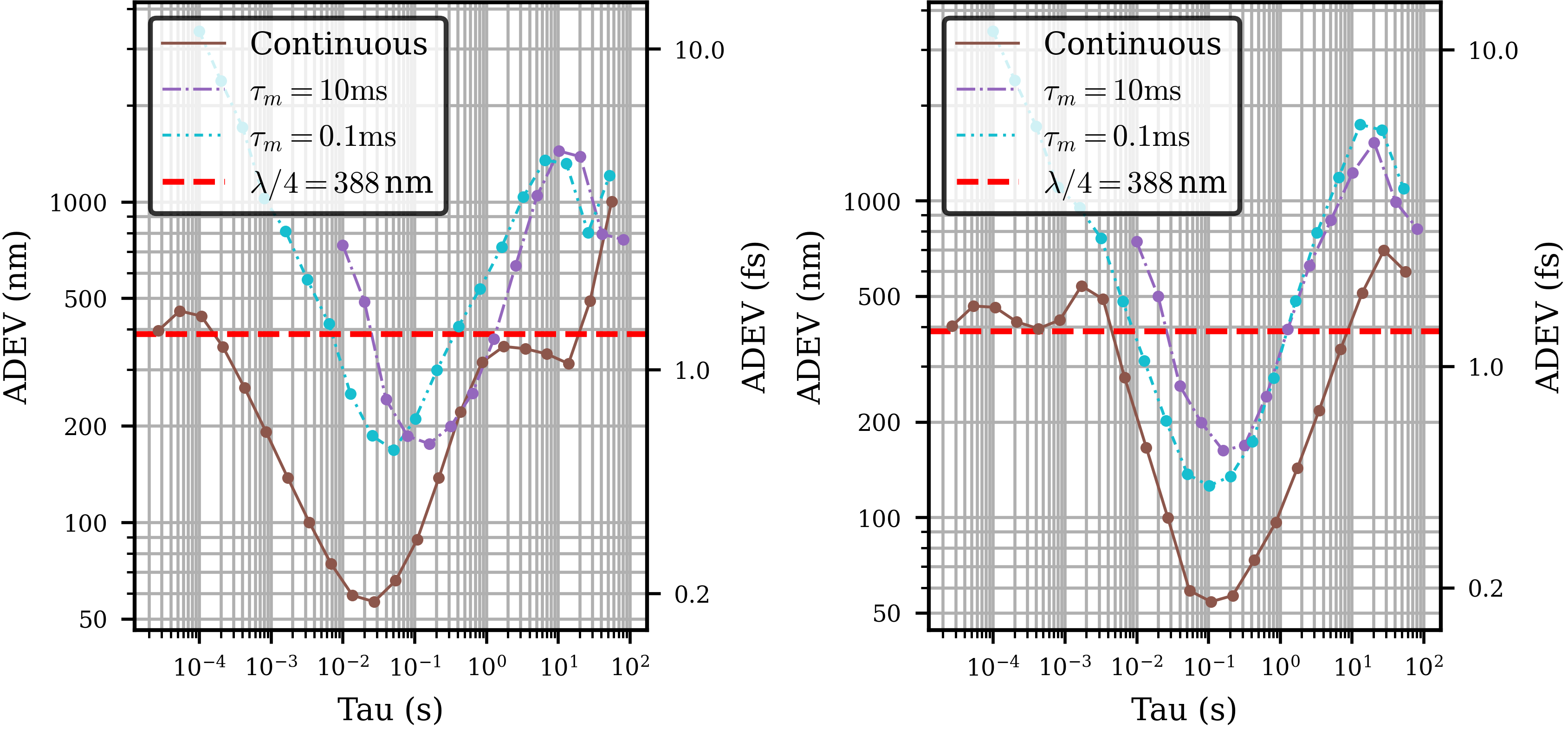}
  \caption{Allan deviation (ADEV) for the short link. The left panel shows the ADEV for a stationary optical delay line, and the right panel shows the ADEV for the same link with \qty{5}{\milli\meter} of continuous displacement. Both the equivalent free-space range (left y-axis) and delay (right y-axis) are provided. The measurement interval is stepped between \qty{0.1}{\milli\second} (cyan) and \qty{10}{\milli\second} (purple). Also provided for comparison is a continuous range measurement (brown) with no reset interval, and the optical quarter wavelength (dashed red).}
  \label{fig:results_fiber_adev}
\end{figure}

\begin{figure}
  \centering
  \includegraphics{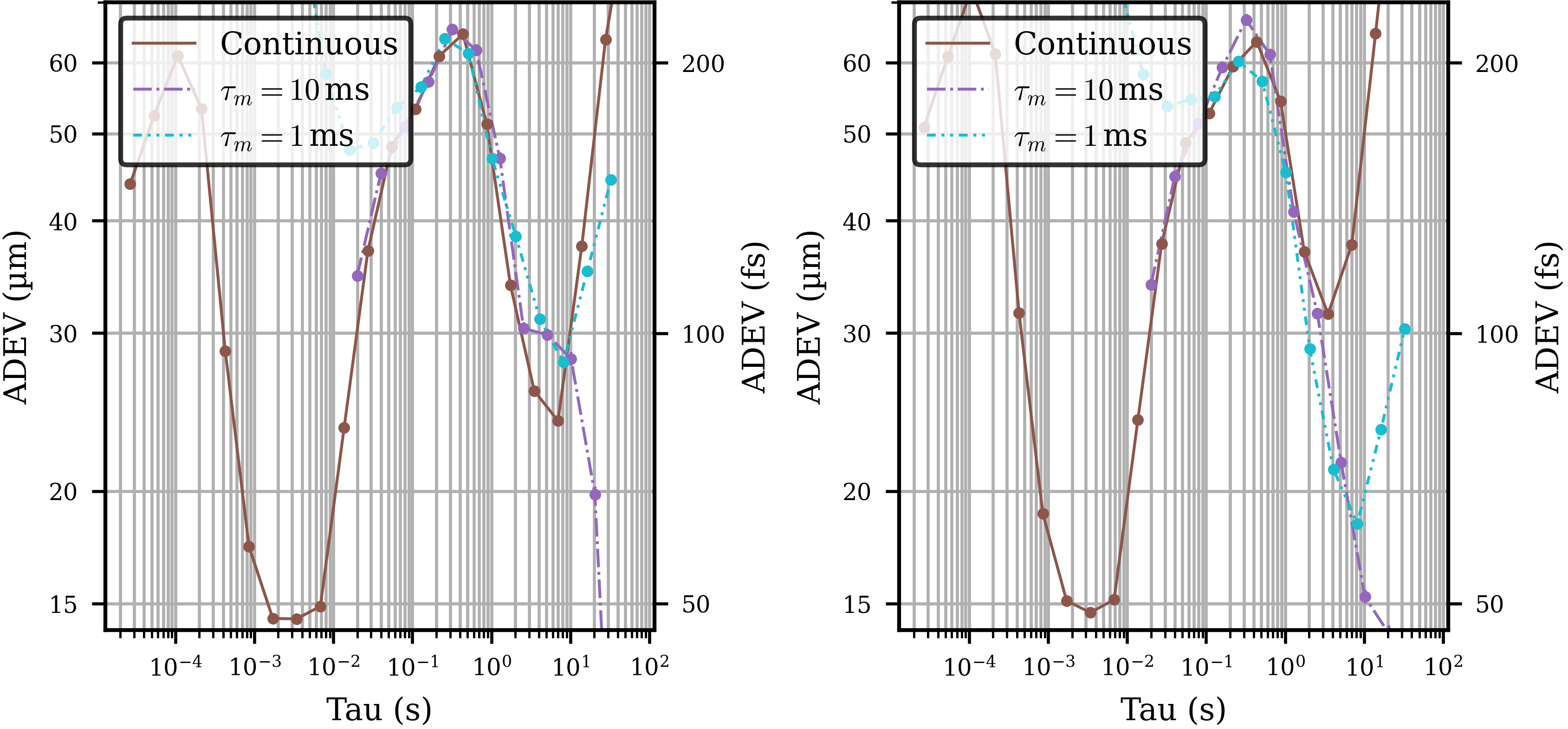}
  \caption{Allan deviation (ADEV) for the \qty{40}{\kilo\meter} link. The left panel shows the ADEV for a stationary optical delay line, and the right panel shows the ADEV for the same link with \qty{5}{\milli\meter} of continuous displacement. Both the equivalent free-space range (left y-axis) and delay (right y-axis) are provided. The measurement interval is stepped between \qty{0.1}{\milli\second} (cyan) and \qty{10}{\milli\second} (purple). Also provided for comparison is a continuous range measurement (brown).}
  \label{fig:results_spool_adev}
\end{figure}

To understand the absolute ranging residuals over different averaging times $\tau$, the ADEV is provided for each experimental result described. The corresponding optical delay for each ADEV is also calculated. This provides the time averaged uncertainty between our absolute range estimate and the carrier displacement for the short link (Fig. \ref{fig:results_fiber_adev}) and the \qty{40}{\kilo\meter} link (Fig. \ref{fig:results_spool_adev}). For the short link, we provide three datasets. The first is a continuous phase measurement, with no PLL resets, shown in brown. The second (purple) and third (cyan) are for PLL measurement intervals of $\tau_m=\qty{10}{\milli\second}$ and $\tau_m=\qty{0.1}{\milli\second}$ respectively. Also provided is the quarter optical wavelength shown in dashed red, which is the required precision to resolve the absolute carrier wavelength. For the \qty{40}{\kilo\meter} link, measurement intervals of \qty{10}{\milli\second} and \qty{1}{\milli\second} are used. These measurements are taken for both the stationary optical delay line (left panel) and moving optical delay line (right panel) in both Figures.

For the short link, we observe a precision of at best \qty{60}{\nano\meter} (\qty{0.2}{\femto\second}), when no wavelength stepping is performed, for both the stationary and moving optical delay line measurements. When wavelength stepping is enabled, this precision worsens to roughly \qty{200}{\nano\meter} (\qty{0.7}{\femto\second}). This is attributed to a decrease in RF power when the RFDC is continuously resetting the NCO. As mentioned previously the short term variations are set by the shot noise limit in Eqn.~\ref{eqn:shot_noise}. For averaging times past $\tau=\qty{1d-1}{\second}$, the precision worsens for each measurement. However, each measurement is capable of handing over to the carrier. The long-term behavior of each measurement is consistent between the continuous phase runs and the stepped phase runs.

The ADEV for the \qty{40}{\kilo\meter} link across each data run remains consistent between $\tau=\qty{1d-2}{\second}$ and \qty{1}{\second}, with a best obtained residual uncertainty of \qty{15}{\micro\meter} (\qty{50}{\femto\second}), or a fractional uncertainty of $\qty{15}{\micro\meter}/\qty{60}{\kilo\meter}\approx\qty{2d-10}{}$. The precision worsens past $\tau=\qty{1d-2}{\second}$ to \qty{60}{\micro\meter} (\qty{200}{\femto\second}). The microwave phase noise previously discussed currently limits the best obtainable precision during short time scales. 
\section{Summary and Conclusion}

In summary, we have demonstrated an absolute synthetic wavelength ranging scheme that takes advantage of a radio-frequency SoC. When compared to the optical carrier displacement, our synthetic range estimate has a best uncertainty less than \qty{60}{\nano\meter} (\qty{0.2}{\femto\second}) at a \qty{18}{\meter} baseline, and a best uncertainty of \qty{15}{\micro\meter} (\qty{50}{\femto\second}) at a \qty{60}{\kilo\meter} (\qty{40}{\kilo\meter} fiber) baseline. We have demonstrated the ability to track physical link variations on the micrometer level using a free-space optical delay line. 

Our scheme does not strictly require careful compensation of polarization or dispersion over long fiber links, which is an advantage when compared to other comb based techniques relying on pulse-to-pulse interference. However, the larger noise floor on the \qty{40}{\kilo\meter} fiber link prevented a ranging precision adequate to resolve the quarter optical wavelength. The higher noise floor is attributed to the phase noise of the microwave tones driving the EOMs, which can be addressed in future work by referencing the RFSoC to a stable microwave reference.

Regardless, this system can still provide valuable ancillary information to optical ranging or optical time and frequency comparison, as it can constrain the optical carrier phase estimates, and in some configurations resolve the absolute optical wavelength. 

\begin{backmatter}
\bmsection{Funding} S.M.P.M is supported by Australian Government Research Training Program Scholarships and top-up scholarships funded by the Government of Western Australia. This work was supported by the Australian Space Agency’s Moon to Mars Demonstrator Mission program with additional funding from the Government of Western Australia and the University of Western Australia.
\vspace{-5pt}
\bmsection{Disclosures} The authors declare no conflicts of interest.
\vspace{-5pt}
\bmsection{Data Availability Statement} Data underlying the results presented in this paper are not publicly available at this time but may be obtained from the authors upon reasonable request.
\vspace{-5pt}
\end{backmatter}

\bibstyle{opticajnl}
\bibliography{biblio.bib}

\end{document}